\documentclass{article}
\usepackage{frascatiphys,here,graphicx,subfigure}

\pdfoutput=1

\begin{document}
\title{Hadronic Physics at CLEO-c}
\author{
M.~R.~Shepherd    \\
{\em Indiana University, Bloomington, Indiana 47405}
}
\maketitle
\baselineskip=11.6pt
\begin{abstract}
The charmonium system provides an opportunity to explore a wide variety of topics in hadronic physics.  Studies of the properties of and transitions among $c\bar{c}$ states yield insight into relativistic and non-perturbative QCD effects.  At the same time, studies of the decays of charmonium states are a window into gluon dynamics and the role of glueball mixing in the production of light quark states.  A collection of preliminary results utilizing the full CLEO-c $\psi(2S)$ data sample is presented including two-body branching fractions of $\chi_{cJ}$ decays, a precision measurement of the $h_c$ mass, and results on the hindered M1 transition:  $\psi(2S)\to\gamma\eta_c$.
\end{abstract}
\baselineskip=14pt
\section{The CLEO-c program}
The CLEO-c physics program is driven by the wide variety of physics accessible in the charmonium region.  Results from large samples of $D$ and $D_s$ decays such as precision measurements of branching fractions and form factors have direct implication on the global heavy-flavor physics program and searches for new particles or interactions outside of the Standard Model.  The CLEO-c program, with the world's largest sample of $\psi(2S)$ decays, is also dedicated to exploring QCD and hadronic physics through studies of the spectrum and decay of $c\bar{c}$ states below open-charm threshold -- this physics topic is the focus of what follows.

The CLEO-c detector, from the beam axis outward, consists of a six-layer stereo inner drift chamber, a 47-layer main drift chamber, a ring-imgaging \v Cerenkov detector (RICH), and a CsI(Tl) crystal calorimeter.\footnote{Outside of the magnet flux return is a series of muon chambers; however, with a detection threshold of approximately 1.2~GeV/$c$, the chambers, while ideal for $B$ physics, are only marginally useful for charm physics.}  The entire detector is immersed in 1-T solenoidal magnetic field.  The two drift chambers provide a momentum resolution of $\approx0.6\%$ for tracks traversing all layers of the chambers.  Photon reconstruction with the 7784-crystal calorimeter is achieved with an energy resolution of $\approx5\%$ at 100 MeV and $\approx2\%$ at 1 GeV.  The detector covers roughly 93\% of the full solid angle.  The results that follow are derived from a sample of approximately 25 million $\psi(2S)$ decays produced at rest in the lab by symmetric $e^+e^-$ collisions in the Cornell Electron Storage Ring (CESR).
\section{Two-body \boldmath $\chi_{cJ}$ decays}
The three $\chi_{cJ}(1P)$ states are readily produced in electromagnetic (E1) transitions from the $\psi(2S)$ and provide a venue for the study of a wide variety of QCD phenomena.  Each of the analyses below rely on full reconstruction of the entire decay chain:  $\psi(2S)\to\gamma\chi_{cJ}$, $\chi_{cJ}\to X$.  The initial four-momentum of the $\psi(2S)$ is well known from the beam kinematics; therefore, the experimental resolution can be enhanced by doing a four-contraint kinematic fit of the decay products to the $\psi(2S)$ four-momentum hypothesis.

\subsection{$\chi_{c0,2}\to\gamma \gamma$}
The two-photon decays of the $\chi_{c0,2}$ states are an ideal place to study relativistic and radiative corrections to QCD in the charmonium system.  To first order these decays are purely QED.  Particular interest is in the ratio $\mathcal{R}\equiv \Gamma(\chi_{c2}\to\gamma\gamma)/\Gamma(\chi_{c0}\to\gamma\gamma)$, which can be calculated simply at first order as $\mathcal{R}=4/15$\cite{Barbieri:1975am}.  Deviations from this lowest order calculation result from radiative corrections and relativistic effects; therefore, precise experimental determination of this ratio is important for validating theoretical calculations that consider these effects.

\begin{figure}
\begin{center}
\includegraphics[scale=0.22]{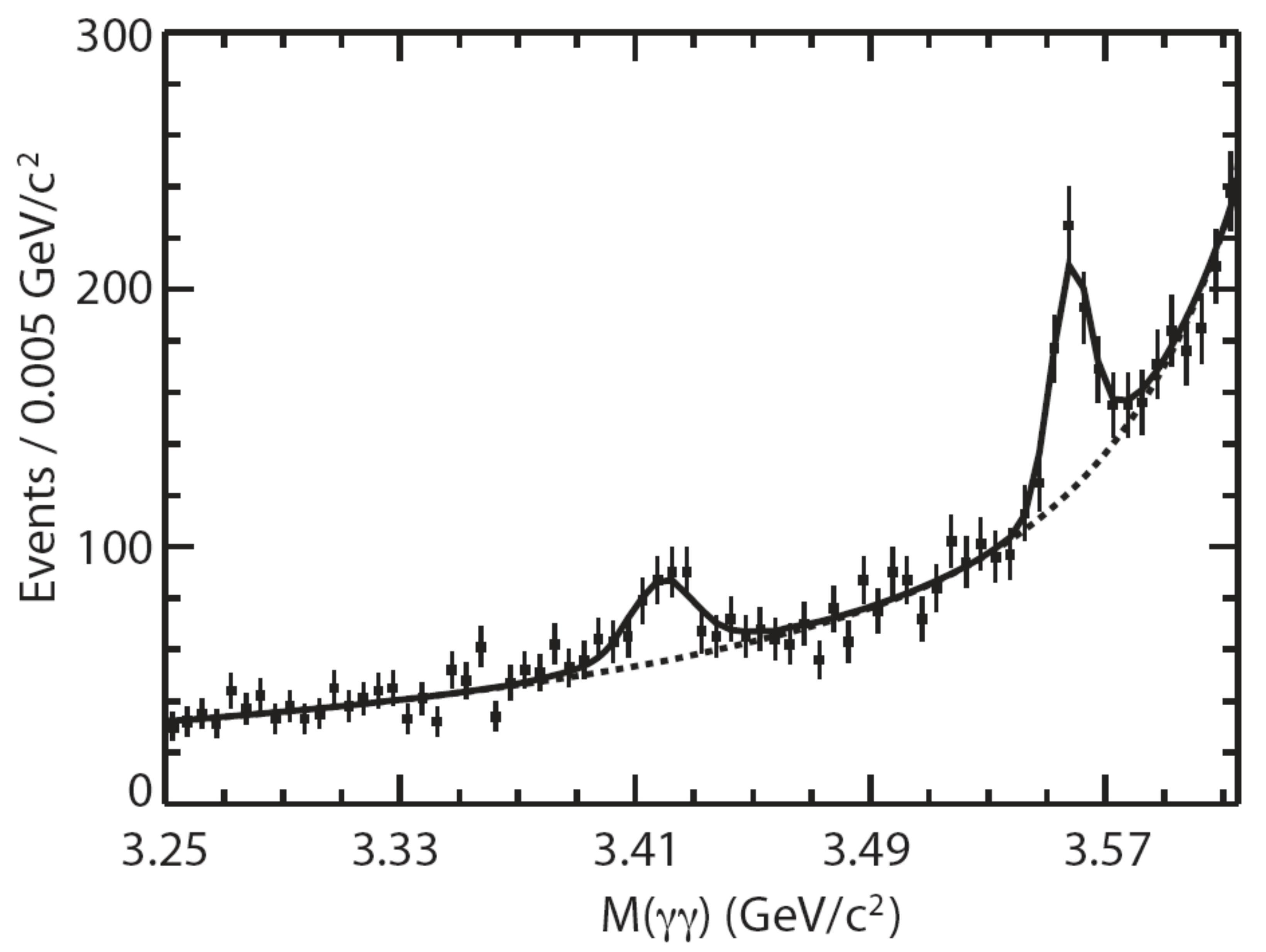}
\caption{Two-photon invariant mass for $\chi_c$ candidates.  The $\chi_{c0}$ and $\chi_{c2}$ peaks are visible.  The fit to the spectrum (background contribution) is shown by the solid (dashed) line.}
\label{fig:chic_2gam}
\end{center}
\end{figure}

To measure the two-photon widths, events are selected that contain three showers greater than 70 MeV in energy and with $|\cos \theta| < 0.75$.  Events with charged particles are vetoed.  The signal shapes and efficiencies are derived from a signal Monte Carlo (MC) sample that was generated using the nominal masses and widths of the $\chi_c$ states\cite{pdg}.  The angular distributions were modeled as a pure E1 transition from the (beam-axis polarized) $\psi(2S)$ to the $\chi_c$ states.  The decay of the $\chi_{c2}$ state was modeled as a pure helicity-two decay\cite{Li:1990sx}.\footnote{As a check on the systematic uncertainty due to this assumption up to an 8\% helicity-zero component was included, motivated by experimental limits in $a_2$ decay\cite{Behrend:1989rc}.}  QED-dominated background shapes are obtained from analyzing data off of the $\psi(2S)$ resonance at $\sqrt{s}=3.671$ GeV and $\sqrt{s}=3.772$ GeV.

Figure~\ref{fig:chic_2gam} shows the fit to the two-photon invariant mass in data.  A total of $212\pm31$ and $335\pm35$ events are observed for $\chi_{c0}\to\gamma\gamma$ and $\chi_{c2}\to\gamma\gamma$ respectively.  A summary of the preliminary results appears in Table~\ref{tab:chic_2gam_tab}.  External measurements for $\mathcal{B}(\psi(2S)\to\gamma\chi_{cJ})$ and $\Gamma_{\mathrm{tot}}(\chi_{cJ})$ are needed to obtain two-photon branching fractions and partial widths.  For these, the PDG average is used\cite{pdg}, and errors due to these external inputs appear as a third, separate systematic error in the table.  Dominant experimental systematic errors are due to the fitting method and determination of the signal efficiency.

\begin{table}
\begin{center}
\caption{Various measured parameters for the decays $\chi_{c0,2}\to\gamma\gamma$.  The errors are, in order, statististical, systematic, and, where applicable, that due to uncertainty in $\mathcal{B}(\psi(2S)\to\gamma\chi_{c0,2})$ and $\Gamma_{\mathrm{tot}}(\chi_{c0,2})$. {\em All results are preliminary.}}
\label{tab:chic_2gam_tab}
\begin{tabular}{cc} \hline\hline
Parameter & This Measurement \\ \hline
$\mathcal{B}(\psi(2S)\to\gamma\chi_{c0})\times\mathcal{B}(\chi_{c0}\to\gamma\gamma)\times 10^5$ & $2.32\pm0.33\pm0.15$ \\
$\mathcal{B}(\psi(2S)\to\gamma\chi_{c2})\times\mathcal{B}(\chi_{c2}\to\gamma\gamma)\times 10^5$ & $2.82\pm0.29\pm0.21$ \\
$\mathcal{B}(\chi_{c0}\to\gamma\gamma)\times10^4$ & $2.52\pm0.36\pm0.16\pm0.11$ \\
$\mathcal{B}(\chi_{c2}\to\gamma\gamma)\times10^4$ & $3.20\pm0.33\pm0.24\pm0.18$ \\
$\Gamma(\chi_{c0}\to\gamma\gamma)$ [keV] & $2.65\pm0.38\pm0.17\pm0.25$ \\
$\Gamma(\chi_{c2}\to\gamma\gamma)$ [keV] & $0.62\pm0.07\pm0.05\pm0.06$ \\
$\mathcal{R}\equiv\Gamma(\chi_{c2}\to\gamma\gamma)/\Gamma(\chi_{c0}\to\gamma\gamma)$ & $0.235 \pm 0.042\pm0.005\pm0.030$ \\ \hline\hline
\end{tabular}
\end{center}
\end{table}

The obtained preliminary result $\mathcal{R}=0.24\pm0.05$ is the most precise single measurement to date; however, it is not yet precise enough to clearly validate any one particular approach for handling both radiative and relativistic corrections in the charmonium system.  When combined with the PDG value of $\mathcal{R}=0.18\pm0.03$ one obtains $\mathcal{R}=0.20\pm0.03$, a value that only marginally disagrees with the zeroth-order prediction of $\mathcal{R}=4/15=0.27$ and motivates more careful experimental and theoretical scrutiny.

\subsection{$\chi_{cJ}\to\eta^{(\prime)}\eta^{(\prime)}$}

Hadronic decays of $\chi_c$ states, like $J/\psi$, proceed dominantly through annihilation into gluons and provide an ideal environment to try to understand gluon dynamics and glueball production.  Hadronic $J/\psi$ decay has received much attention in this regard due to the interesting series of results from BES\cite{besglue} that study production of scalar $f_0$ resonances against flavor-tag $\omega$ and $\phi$ states in $J/\psi$ decay.  Close and Zhao interpret the results, which appear to be suggestive of large OZI rule violating effects, as a signature for scalar glueball mixing amongst the $f_0$ states in the 1.5 GeV/$c^2$ region.  Following up on this work, Zhao proposes a factorization scheme in which one can coherently analyze the partial widths of various two-body $\chi_c$ decays\cite{zhao} in terms of singly and doubly OZI suppressed components, where a large doubly OZI suppressed component could also be indicative of strong glueball mixing.  A coherent study of $\chi_{cJ}\to\eta^{(\prime)}\eta^{(\prime)}$ provides a testing ground for this production factorization model.  In addition Thomas notes\cite{Thomas:2007uy} that these decays provide a mechanism to explore the gluonic component of the $\eta^\prime$.

Like the two-photon decays, analysis of $\chi_{cJ}\to\eta^{(\prime)}\eta^{(\prime)}$ relies on the reconstruction and kinematic fit of the entire event.  This analysis is an update of a previous CLEO-c analysis that utilized a subset of the data\cite{Adams:2006na}.  The $\eta$ decay candidates are detected in the modes $\gamma\gamma$, $\pi^+\pi^-\pi^0$, and $\gamma\pi^+\pi^-$, while the $\eta^\prime$ candidates are reconstructed in $\gamma\pi^+\pi^-$ and $\eta\pi^+\pi^-$ modes.  After the kinematic fit, the two body invariant mass distribution is plotted and signals are extracted by fitting the peaks to a Breit-Wigner with widths fixed by the PDG values\cite{pdg} convoluted with a MC-determined Gaussian resolution.  In calculating experimental efficiencies, it is assumed that $\psi(2S)\to\gamma\chi_{c0,2}$ is a pure E1 transition.  Branching fractions are summarized in Table~\ref{tab:chiceta}.

\begin{table}
\begin{center}
\caption{Measured branching fractions and 90\% confidence level upper limits for $\chi_{c0,2}\to\eta^{(\prime)}\eta^{(\prime)}$.  Where applicable, the errors are statistical, systematic, and due to $\mathcal{B}(\psi(2S)\to\gamma\chi_{cJ})$ respectively. {\em All results are preliminary.}}
\label{tab:chiceta}
\begin{tabular}{l|cc}\hline\hline
$X$ & $\mathcal{B}(\chi_{c0}\to X)\times10^3$ & $\mathcal{B}(\chi_{c2}\to X)\times10^3$ \\ \hline
$\eta\eta$ & $3.18\pm0.13\pm0.18\pm0.16$ & $0.51\pm0.05\pm0.03\pm0.03$ \\
$\eta\eta^\prime$ & $<0.25$ & $<0.05$ \\
$\eta^\prime\eta^\prime$ & $2.12\pm0.13\pm0.11\pm0.11$ &  $0.06\pm0.03\pm0.004\pm0.004$ \\
 & & $(<0.10)$ \\ \hline\hline
\end{tabular}
\end{center}
\end{table}

\begin{figure}
\begin{center}
\includegraphics[scale=0.29]{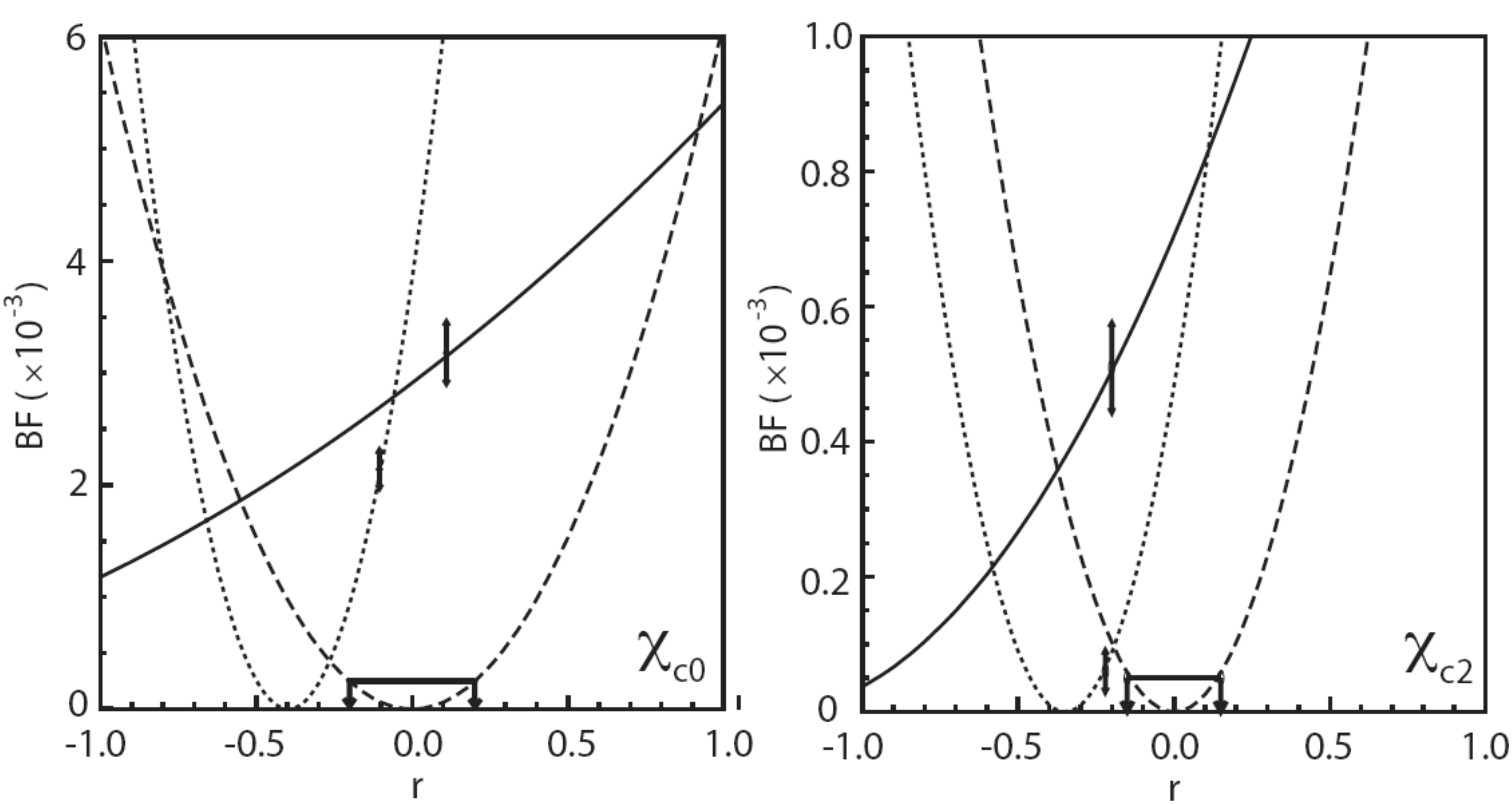}
\caption{Predictions for the branching fraction of $\chi_{c0,2}$ to $\eta\eta$ (solid), $\eta\eta^\prime$ (dashed), and $\eta^\prime\eta^\prime$ (dotted) as a function the model parameter $r$, which is the ratio of doubly to singly OZI suppressed decays.  Experimental measurements presented in this work are indicated by the arrows.}
\label{fig:etaeta}
\end{center}
\end{figure}

The Zhao model casts the branching fractions as a function of $r$, a parameter that is equal to the ratio of the strengths of doubly OZI to singly OZI suppressed decays.  Figure~\ref{fig:etaeta} shows the predicted branching fractions for the $\eta\eta$, $\eta\eta^\prime$, and $\eta^\prime\eta^\prime$ final states as a function of $r$ for both the $\chi_{c0}$ and $\chi_{c2}$.  One notes that for both the $\chi_{c0}$ and $\chi_{c2}$ all measured values or limits are consistent with the same value of $r$ lending support for the validity of the model.  One also notes that the value of $r$ is close to zero which indicates a relatively small component of doubly OZI suppressed production, consistent with what is commonly accepted as small glueball mixing amongst the isoscalar pseudoscalar mesons.  Applications of this technique to the scalar meson sector are underway at CLEO-c and BES.

\subsection{Other two-body decays}

In addition to probing the role of glueball mixing in production, two-body decays of $\chi_c$ states provide an opportunity to explore the role of the color octet mechanism (COM) in $P$-wave charmonia. The COM was proposed\cite{Wong:1999hc} to explain the apparent deficit in theoretically predicated decay rates for nucleon-antineucleon pairs based on the color singlet model, and therefore motivates a new series of precision measurements of two-body $P$-wave charmonia decay.

The analysis of these decays proceeds in the same fashion as those mentioned above, namely exploiting the power of the full event kinematic fit to improve resolution and reduce background.  The final states reconstructed are listed in Tables~\ref{tab:2mes} and~\ref{tab:2bar}.  The hyperon decays are reconstructed in the following modes:  $\Lambda\to p\pi^-$, $\Sigma^+\to p\pi^0$, $\Sigma^0\to\Lambda\gamma$, $\Xi^-\to\Lambda\pi^-$, and $\Xi^0\to\Lambda\pi^0$.  As with the two-photon decays, the kinematically constrained two-body invariant mass distributions are fit to extract the yield for each of the $\chi_c$ signal peaks.  Experimental efficiency is determined using a MC simulation.  For decays of the $\chi_{c1}$ and $\chi_{c2}$ to two hyperons, the helicity of the final state is unknown and the range of efficiencies for the allowed helicity configurations is used to quantify the systematic error due to this uncertainty.

The results are summarized in Tables~\ref{tab:2mes} and~\ref{tab:2bar} and represent the most precise measurements to date of these two-body branching fractions.  While a detailed comparison of the results the COM predictions\cite{Wong:1999hc} is not possible here, in general, measured branching fractions tend to be higher than those predicted by the COM-motivated predictions suggesting further theoretical understanding of these decays is needed.

\begin{table}
\begin{center}
\caption{Measured branching fractions for various two-meson $\chi_{cJ}$ decays.  The errors are statistical, systematic, and due to $\mathcal{B}(\psi(2S)\to\gamma\chi_{cJ})$ respectively. {\em All results are preliminary.}}
\label{tab:2mes}
\begin{tabular}{l|cc}\hline\hline
$X$ & $\mathcal{B}(\chi_{c0}\to X)\times10^3$ & $\mathcal{B}(\chi_{c2}\to X)\times10^3$ \\ \hline
$\pi^+\pi^-$ & $6.37\pm0.11\pm0.20\pm0.32$ & $1.59\pm0.04\pm0.06\pm0.10$ \\
$\pi^0\pi^0$ & $2.94\pm0.07\pm0.16\pm0.15$ & $0.68\pm0.03\pm0.05\pm0.04$ \\
$K^+K^-$ & $6.47\pm0.11\pm0.29\pm0.32$ & $1.13\pm0.03\pm0.05\pm0.07$\\
$K^0_SK^0_S$ & $3.49\pm0.01\pm0.15\pm0.17$ & $0.53\pm0.03\pm0.02\pm0.03$\\ \hline\hline
\end{tabular}
\end{center}
\end{table}

\begin{table}
\begin{center}
\caption{Measured branching fractions for various two-baryon $\chi_{cJ}$ decays.  The errors are statistical, systematic, and due to $\mathcal{B}(\psi(2S)\to\gamma\chi_{cJ})$ respectively. {\em All results are preliminary.}}
\label{tab:2bar}
\begin{tabular}{l|cr@{\hspace{2pt}$\pm$\hspace{2pt}}lr@{\hspace{2pt}$\pm$\hspace{2pt}}l}\hline\hline
$X$ & $\mathcal{B}(\chi_{c0}\to X)\times 10^5$ & \multicolumn{2}{c}{$\mathcal{B}(\chi_{c1}\to X)\times 10^5$} & \multicolumn{2}{c}{$\mathcal{B}(\chi_{c2}\to X)\times 10^5$} \\ \hline
$p\bar{p}$ & $25.7\pm 1.5\pm1.5\pm1.3$ & $9.0$ & $0.8\pm0.4\pm0.5$ & $7.7$ & $0.8\pm0.4\pm0.5$ \\
$\Lambda\bar{\Lambda}$ & $33.8 \pm 3.6\pm 2.3\pm1.7$ & $11.6$ & $1.8\pm0.7\pm0.7$ & $17.0$ & $2.2\pm1.1\pm1.1$ \\
$\Sigma^0\bar{\Sigma}^0$ & $44.1\pm5.6\pm2.5\pm2.2$ & $2.1$ & $1.4\pm0.2\pm0.1$ & $4.1$ & $1.9\pm0.3\pm0.3$ \\
$\Sigma^+\bar{\Sigma}^-$ & $32.5\pm5.7\pm 4.9\pm1.7$ & $3.3$ & $1.8\pm0.2\pm0.2$ & $3.3$ & $1.9\pm0.4\pm0.2$ \\
$\Xi^{0}\bar{\Xi}^0$ & $33.4\pm7.0\pm3.2\pm1.7$ & $2.5$ & $2.1\pm0.2\pm0.2$ & $4.0$ & $2.4\pm0.4\pm0.3$ \\
$\Xi^{-}\bar{\Xi}^+$ & $51.4\pm6.0\pm3.8\pm2.6$ & $8.6$ & $2.2\pm0.6\pm0.5$ & $14.5$ & $1.9\pm1.0\pm0.9$ \\ \hline\hline
\end{tabular}
\end{center}
\end{table}

\section{The hindered M1 transition:  \boldmath $\psi(2S)\to\gamma\eta_c$}

A clear experimental picture of both the hindered ($\psi(2S)\to\gamma\eta_c$) and allowed ($J/\psi\to\gamma\eta_c$) M1 transitions in charmonium is important for understanding a variety of theoretical and experimental issues.  For example, there is interest in using radiative transitions in charmonium to explore photon couplings to quarks in lattice QCD\cite{Dudek:2006ej}.  Calculating these two rates has been a challenge for quark models\cite{Eichten:2007qx}.  Both rates are key in normalizing exclusive branching fractions of the $\eta_c$; the focus here is on $\mathcal{B}(\psi(2S)\to\gamma\eta_c$).  As will be discussed in detail below, the $\eta_c$ lineshape in this decay appears to be non-trivial, and this complicates the measurement of the rate.

\begin{figure}
\begin{center}
\includegraphics[scale=0.26]{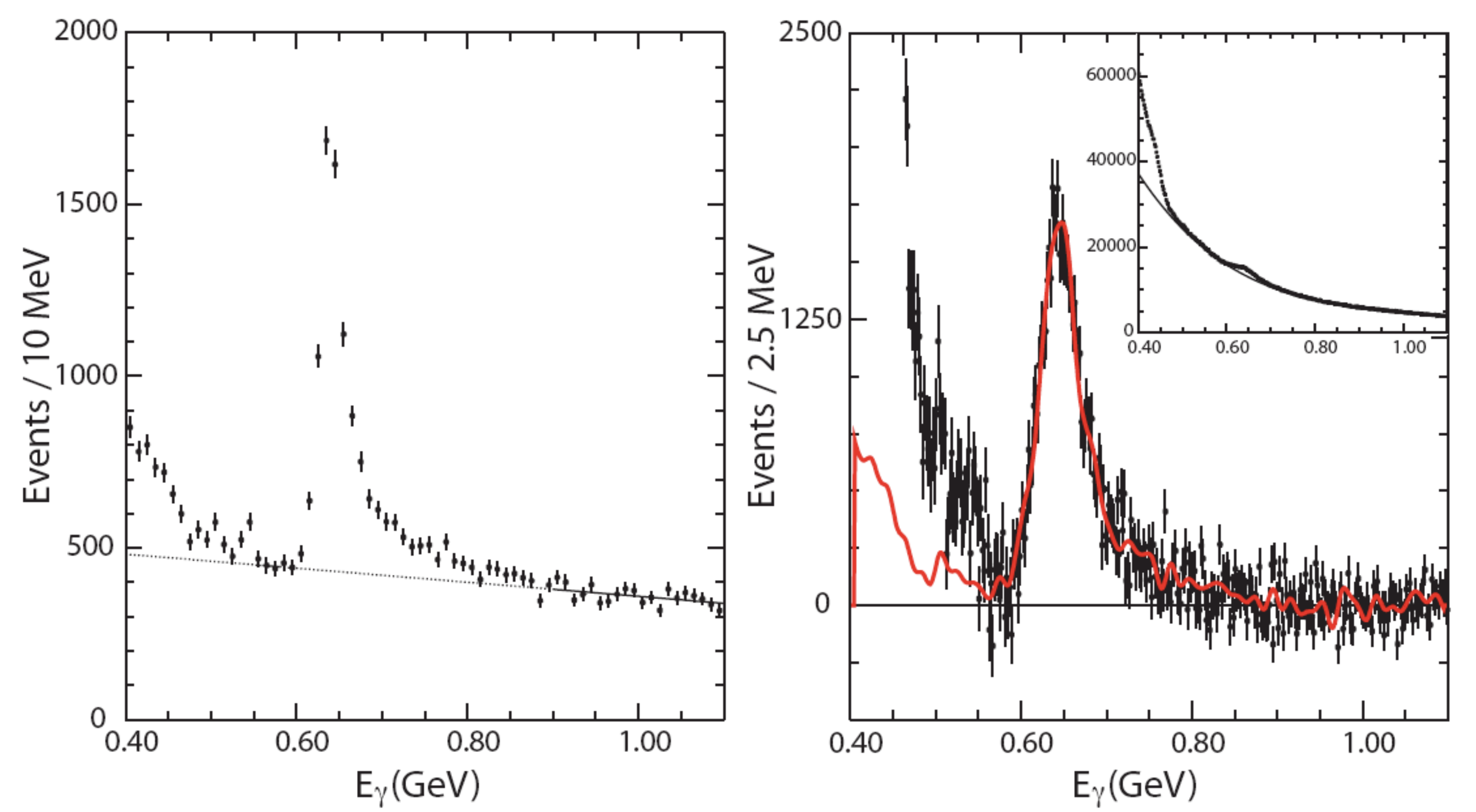}
\caption{Left:  The photon spectrum the process $\psi(2S)\to\gamma\eta_c$ for exclusively reconstructed $\eta_c$ states and after a full event kinematic fit.  The intrinsic lineshape dominates the experimental resolution, which varies from 4-7 MeV depending on $\eta_c$ decay mode.  Right:  The background subtracted inclusive photon spectrum (raw shown in inset).  The exclusive lineshape before the kinematic fit is superimposed as a solid (red) line.}
\label{fig:etac}
\end{center}
\end{figure}

To examine the lineshape in detail, thirteen signal-rich $\eta_c$ decay modes (and charge conjugates) are reconstructed: $2(\pi^+\pi^-)$, $\pi^+\pi^-\pi^0\pi^0$, $3(\pi^+\pi^-)$, $2(\pi^+\pi^-\pi^0)$, $2(K^+K^-)$, $K^+K_S\pi^-$, $K^+K^-\pi^0$, $K^+K^-\pi^+\pi^-$, $K^+K_S\pi^+\pi^-\pi^-$, $K^+K^-\pi^+\pi^-\pi^0$, $K^+K^-2(\pi^+\pi^-)$, $\eta\pi^+\pi^-$, and $\eta2(\pi^+\pi^-)$.  Like the $\chi_{cJ}$ decays, full event reconstruction and kinematic fitting is employed for these candidates.  Figure~\ref{fig:etac} (left) shows the photon spectrum after it has been sharpened by the kinematic fit.  The background is fit to a (MC-motivated) linear function using data in the region $E_\gamma > 900$~MeV and $560<E_\gamma<600$~MeV.  Peaking backgrounds below 560~MeV are due to $h_c\to\gamma\eta_c$, photon cascades from $\psi(2S)$ to $\chi_c$ to $J/\psi$ states,  and $\psi^\prime\to\pi^0J/\psi$, where, for the latter two backgrounds, the two photons merge in the calorimeter.  The signal shows a distinct tail on the high energy side of the photon spectrum.

Modification of the line shape for this transition is expected since the natural width is relatively large and the available phase space grows like $E_\gamma^3$.  In addition the hindered M1 transition has an additional $E_\gamma^2$ term in the matrix element\cite{Brambilla:2005zw} that may enhance the line shape distortion.   These additional line shape modifications, while theoretically motivated, are not constrained well enough to allow a satisfactory fit to the data and lead one to question whether this transition is suitable for extracting the mass and width of the $\eta_c$.

Figure~\ref{fig:etac} (right) shows the exclusive photon spectrum before kinematic fitting (red line) superimposed on the background-subtracted inclusive photon spectrum (points).  The agreement is excellent indicating that the line shape modification is visible also in the raw inclusive photon spectrum.  A variety of techniques are used to extract the yield in the inclusive photon spectrum including using an empirical parametrization of the peak and simply counting events above background.  The 10\% uncertainty in the number of signal events due to uncertainties in background and signal lineshape is the dominant systematic uncertainty in the measurement of $\mathcal{B}(\psi(2S)\to\gamma\eta_c)$.  Our preliminary result is $\mathcal{B}(\psi(2S)\to\gamma\eta_c) = ( 4.02\pm0.11\pm0.52 ) \times10^{-3}$.

\section{The mass of the \boldmath $h_c$ }

The singlet $h_c(^1P_1)$ state was the last of the expected charmonium states below $D\bar{D}$ threshold to be identified\cite{hcdisc}.  There is interest in understanding the hyperfine splittings of the charmonium states as these give insight into the nature of the spin-spin interaction in QCD.  In the limit that the confinement term in the QCD potential carries no spin dependence, one expects non-zero hyperfine splitting for only $L=0$ states, and the mass of the $h_c$ ($L=1$) should be equal to the spin-averaged $\chi_{cJ}$ mass.  Therefore precision measurement of this splitting $\Delta M_{\mathrm{hf}}(1P) = \langle M( 1^3P_J ) \rangle - M( 1^1P_1 )$ provides experimental input on the spin dependence of the $q\bar{q}$ interaction.  The error on $\Delta M_{\mathrm{hf}}$ is currently dominated by error on the mass of the $h_c$; hence, a more precise measurement is desirable.

The $h_c$ is studied in the isospin-violating process $\psi(2S)\to\pi^0 h_c,~h_c\to\gamma\eta_c$.  Two methods are utilized:  one that is inclusive of all $\eta_c$ decay modes and another that reconstructs the $\eta_c$ in a collection of exclusive hadronic modes, which mostly overlap with those noted in the previous section.  Both require a signal $\pi^0$ from the primary transition $\psi(2S)\to\pi^0h_c$ be identified from two-photon candidates within three standard deviations of the $\pi^0$ mass and extract the signal from fits to recoil mass spectrum against this $\pi^0$.  

\begin{figure}
\begin{center}
\includegraphics[scale=.25]{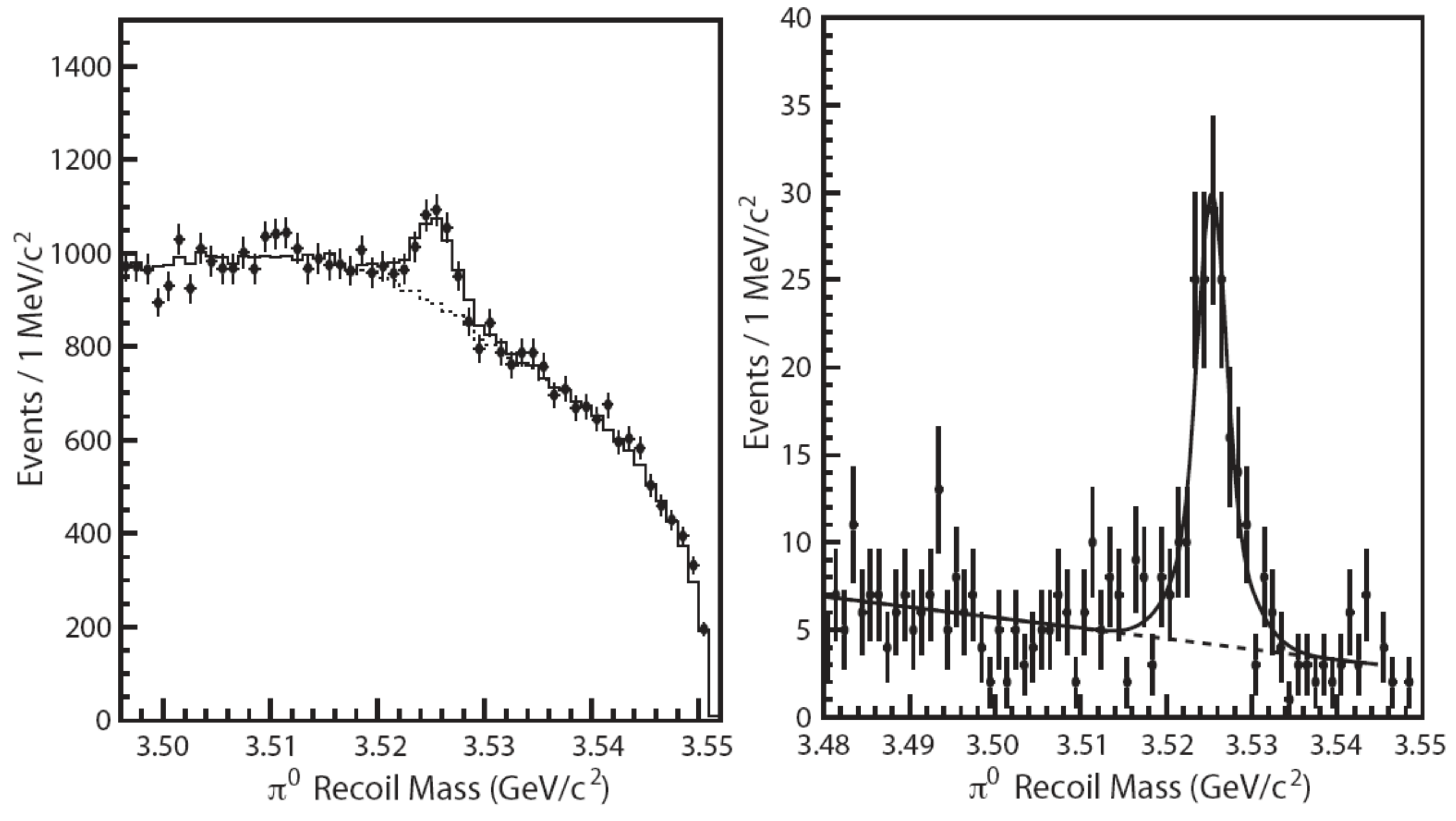}
\caption{Plots of the recoil mass against the $\pi^0$ for inclusive (left) and exclusive (right) $\eta_c$ selection.  The $h_c$ signal is clearly visible in both cases.  Fits to the spectra (background contributions) are shown by the solid (dashed) lines.}
\label{fig:hc}
\end{center}
\end{figure}

The inclusive analysis relies on the identification of a candidate photon for the $h_c\to\gamma\eta_c$ transition with an energy of $503\pm35$~MeV.  Removing this criteria overwhelms the signal with background and allows one to determine the background shape.  Figure~\ref{fig:hc} (left) shows the fitted $\pi^0$ recoil mass spectrum.  The signal shape is Breit-Wigner with width fixed to that of the $\chi_{c1}$ convoluted with a Gaussian resolution function of width 2.5~MeV$/c^2$.  The mass obtained is $3525.35\pm0.24\pm0.21$~MeV$/c^2$.  The angular distribution of the photon in the $h_c\to\gamma\eta_c$ is consistent with that of an $E1$ transition:  $dN/d\cos \theta_\gamma \propto 1 + \alpha \cos^2\theta_\gamma$, where $\alpha=1$.  We obtain $\alpha=1.34\pm0.53$ from the data.

The exclusive analysis, like other exclusive analyses mentioned above, relies on full reconstruction and kinematic fit of the entire decay chain.  The mass of the $\eta_c$ candidate was required to be within 30~MeV$/c^2$ of the nominal $\eta_c$ mass\cite{pdg}.  The $\pi^0$ recoil spectrum from the set of exclusive candidates is shown on the right of Figure~\ref{fig:hc} and is fit to a linear background plus a Breit-Wigner convoluted with a double-Gaussian resolution function obtained from MC.  The mass obtained from the fit is $3525.35\pm0.27\pm0.20$~MeV$/c^2$.

Accounting for statistical correlations between the exclusive and inclusive samples, we obtain the preliminary result:  $M(h_c) = 3525.35\pm0.19\pm0.15$~MeV$/c^2$.  This yields a hyperfine splitting $\Delta M_{\mathrm{hf}}(1P) = -0.05\pm0.19\pm0.16$~MeV/$c^2$, which is remarkably consistent with zero.  However, Richard\cite{Richard:2007wc} cautions against interpreting this result as a lack of spin-spin interactions in the $1P$ multiplet as $M(h_c)$ should really be compared with the spin-averaged $\chi_c$ mass as calculated in the potential model, which is several MeV higher.  

\section{Summary}

The charmonium system provides a rich landscape to study QCD.  We have presented precision measurements for many two-body decays of the $\chi_{cJ}$ states, which have implications in understanding relativistic and radiative corrections in the charmonium system, the role of the color octet mechanism in $P$-wave decay, and glueball mixing amongst the light scalar mesons.  The hindered M1 transition, $\psi(2S)\to\gamma\eta_c$, exhibits a non-trivial lineshape that is necessary to understand theoretically if precision experimental measurements for the partial width for this decay and the mass and width of the $\eta_c$ are to be obtained.  Finally we presented a new precision measurement of the mass of the $h_c$ that is naively consistent with zero hyperfine splitting in the $1P$ multiplet of charmonium.  

Analysis of the large sample of $\psi(2S)$ collected with the CLEO-c detector continues, and more exciting hadronic physics results are expected in the coming year.  I would like to acknowledge the work of my CLEO colleagues on these analyses.  I would also like to thank the \textsc{Hadron 07} organizers for providing a wonderful venue to present these results.

\end{document}